\documentclass[letterpaper]{article}
\usepackage{iccc}

\usepackage{times}
\usepackage{helvet}
\usepackage{courier}
\usepackage{graphicx}
\usepackage{booktabs}
\usepackage{array} 
\usepackage{amsmath}
\usepackage{algpseudocode}

\newcommand{\MP}{\emph{Mimetic Poet}}

\title{Mimetic Poet}
\author{Jon McCormack, Elliott Wilson, Nina Rajcic and Maria Teresa Llano\\
SensiLab\\
Monash University\\
Caulfield East, Victoria, AU\\
Jon.McCormack@monash.edu \hspace{0.1cm} Elliott.Wilson@monash.edu \hspace{0.1cm} Nina.Rajcic@monash.edu \hspace{0.1cm} Teresa.Llano@monash.edu\\
}

\begin{document} 
\maketitle
\begin{abstract}
This paper presents the design and initial assessment of a novel device that uses generative AI to facilitate creative ideation, inspiration, and reflective thought. Inspired by magnetic poetry, which was originally designed to help overcome writer's block, the device allows participants to compose short poetic texts from a limited vocabulary by physically placing words on the device's surface. 
Upon composing the text, the system employs a large language model (LLM) to generate a response, displayed on an e-ink screen. We explored various strategies for internally sequencing prompts to foster creative thinking, including analogy, allegorical interpretations, and ideation. We installed the device in our research laboratory for two weeks and held a focus group at the conclusion to evaluate the design. The design choice to limit interactions with the LLM to poetic text, coupled with the tactile experience of assembling the poem, fostered a deeper and more enjoyable engagement with the LLM compared to traditional chatbot or screen-based interactions. This approach gives users the opportunity to reflect on the AI-generated responses in a manner conducive to creative thought.
\end{abstract}

\section{Introduction}
\label{s:introduction}
The advent of large generative AI models, such as transformers, heralds a new era in language-based interaction with machines. However, despite much potential, the majority of interactions with generative AI systems, such as Large Language Models (LLMs) is via simple ``chat'' interfaces and the form of that interaction is commonly in a question and answer format. This form of interaction reinforces the notion that the machine interlocutor is an ``intelligent'' but subservient entity, task-focused, willing to please and able to answer questions put to it, but unable to initiate its own dialogue because it lacks intention \cite{McCormack2024b}. Moreover, any dialogue-based interaction via screens does not consider implicit context, something that is particularly important in creative applications and in forming aesthetic judgements \cite{Leder2014}.

In this paper we propose an alternative way to engage with LLMs, building a novel form of interface designed to support ideation and creative thinking, either for individuals or groups. Called the \MP, the device augments traditional magnetic poetry, using it as the mechanism for communicating with an LLM in an interactive dialogue over extended periods of time (days or weeks). Through the use of constraints and physical interactions, the device promotes a slower form of interaction with LLMs, allowing space for contemplation and thought beyond conventional dialogue-based screen interfaces.

In the sections that follow we first introduce some background rationale for building this device and for the mode of interaction it uses. This rationale draws together theories of intention in writing and in design as a basis for interacting with a language-based AI model, which somewhat paradoxically, has no authorial intention. This turns out to be an advantage as it enables us to place both human and AI on a more equal footing in terms of intention.

Next we describe the system design and its technical operation in detail, including both the physical design and internal prompt chaining with the LLM (OpenAI's GPT-4 API). To evaluate our system we installed it in our research lab for a period of two weeks and asked members of the lab to use it whenever they felt like it. At the conclusion of this period we held a focus group to draw some preliminary findings on the system's efficacy as both an interface and as a way of promoting ideation and creative thought.

Lastly we reflect on the current limitations of the device and briefly discuss possibilities for improvement, along with some more general observations on people's perception of current generative AI models.

\section{Background and Related Work}

\subsection{The Intentional Fallacy}
\label{ss:intentional_fallacy}

There's a general agreement in literary theory that the meaning of a text does not reside exclusively with the author. Known as the ``intentional fallacy'' \cite{Wimsatt1946}, it accounts for the idea that we may not know what the author of a text intended, and cannot ask them if, for example, they are no longer living. Additionally, the interpretation of meaning in a text is not exclusively determined by the author, where unintended meaning may arise due to historical or cultural context, or the author might deliberately lie or try to obscure the meaning. The philosopher Don Ihde sees parallels in technological design, describing what he calls the ``designer fallacy'' in thinking that ``a designer can design into a technology, its purposes and uses'' \cite{Ihde2008}. Ihde is a leading proponent of \emph{postphenomenology} \cite{Ihde1993}, a methodological tool widely used in human-computer interaction (HCI) design to analyse the relationships between humans and technology, in particular how technology mediates our view of the world and our actions in it. Postphenomenology sits in contrast to more traditional approaches to interface design, such as ``user-centered design'' \cite{Norman1986}, which generally limits design considerations to the direct interactions between human and machine.

This dualism of intentional fallacy in text and designer fallacy in interface design is something we weave together in this work, which explores relationships people might have with the emerging new technologies of generative artificial intelligence. We explore design possibilities beyond conventional forms of interaction with LLMs, such as ``chatbot'' interfaces popularised by systems such as OpenAI's ChatGPT, instead designing speculative possibilities for what human-AI relationships \emph{could be} \cite{rajcic2023message}. The value of this form of design extends beyond functionality or aesthetics. By challenging assumptions and conceptions about the role that objects and technology play in our lives, this form of speculative design---as process and way of thinking---can serve as a means of stimulating different ways of ``speculat[ing] about possible futures; and as a catalyst for change.'' \cite[p.33]{Dunne2013}.

If we accept that the technological designer's intention does not fully determine how people may use their design, nor its broader socio-cultural effects, then we are free to consider the possibilities a design might facilitate rather than mandate (e.g.~rather than asking \emph{``what is its function or purpose?''} we consider \emph{``what might it make possible?''}). In speculative design we design possible futures to better understand and evaluate the implications of new technologies before they become embedded in society; to enact change rather than conforming to the status quo \cite{Dunne2013}. This becomes particularly important with the rapid, and largely unregulated, rise of generative AI and its ethical, social and cultural impact on human creativity \cite{rajcic2024towards}.

The intentional fallacy also finds new meaning in generative AI systems, in particular LLMs which can generate coherent text with apparent intention and meaning. However, an LLM does not have a discernible authorial intent, rather each output is a statistical accumulation of individual authorial intent, from those whose work was included in the training data\footnote{The exception being in cases where LLMs can be coaxed into verbatim repeating of their training data \cite{carlini2021extracting,nasr2023scalable}.}. Moreover, the models are prone to ``hallucinations'' where they produce factually incorrect or semantically misleading output. If the intention doesn't come from the LLM, and what the model produces may be a hallucination, then there may be situations where the reader is freer to make their own interpretation. This includes scenarios where reliance on factual or explanatory information is not essential, such as divergent creative thinking, the use of analogy or metaphor in activities such as brainstorming or creative ideation.

\subsection{Creativity Support Tools}
\label{ss:cst}

The study of creativity has been enriched by the development of theories such as divergent thinking \cite{mccrae1987creativity}, which emphasizes the generation of multiple solutions to a given problem. This concept, introduced by Guilford in the 1950s, underscores the importance of thinking in varied and unique directions as a hallmark of creativity. Additionally, the theory of intrinsic motivation \cite{hennessey1998reality} highlights the role of self-motivation in fostering creativity, suggesting that creative endeavors are most fruitful when driven by genuine interest and satisfaction in the work itself.

On the other hand, Creativity Support Tools (CSTs) are generally designed to aid individuals in furthering their creative faculties by providing resources, inspiration, and technological scaffolding to navigate the creative process. These tools range from simple brainstorming tools that facilitate idea generation through divergent thinking, such as Brian Eno's \emph{Oblique Strategies} \cite[Ch.1]{Harford2016}, and towards more complex systems that integrate technology into the process. However, ways of understanding how generative AI systems might support creativity via creative strategies is relatively underdeveloped \cite{Koch2019,Koch2020,Verheijden2023}.

In the evolving landscape of computational creativity, generative text has been explored through various developments of algorithmic sophistication. Early research explored the modeling of emotion within algorithmically generated poetry \cite{misztal2014poetry}, the constraining of poetic structure of generated poems \cite{colton2012full}, and poetry as responding to input images \cite{loller2018deep}. 

As technology advanced, research shifted towards interactive and collaborative systems \cite{oliveira2019co}. Research also looked into the development of generative poetry as a Creativity Support Tool (CST) for creative writers \cite{booten2021poetry}, as well as for creativity in general \cite{kantosalo2014isolation}. This evolution reflects a growing interest in not only automating the generation of poetic content but also in developing systems that can engage with human users in a more meaningful, collaborative manner.

Researchers have furthermore explored the embedding of generative poetry into physical interfaces \cite{rajcic2020mirror,rajcic2023message,stiles_technelegy_2022} introducing an innovative dimension to the interaction between humans and computational creative systems. Such approaches not only challenge the traditional boundaries of poetic expression but also invite users to engage with poetry in more tangible and immersive ways, increasing accessibility of generative poetry, by allowing engagement without the need for conventional screen interfaces.

\subsection{Found Poetry}
\label{ss:found_poetry}

Magnetic Poetry, a literary phenomenon characterized by the arrangement of individual words on magnetic surfaces to create poetry, emerged as a modern iteration of the found poetry technique. The genesis of Magnetic Poetry can be traced back to 1993, when songwriter Dave Kapell, seeking to overcome writer's block, developed the concept by scattering words on his refrigerator, allowing for spontaneous and serendipitous writing. This innovation offered a tangible medium for creative writing while making it fun and accessible to a wider audience without the prerequisite of formal literary training.

The technique of ``found'' poetry has a rich historical lineage extending into the early 20th century. Found poetry re-purposes existing texts, extracting and rearranging words and phrases to form new meanings and poetic expressions \cite{barda2019techniques}. This method challenges traditional notions of authorship and creativity, suggesting that art can emerge from the re-contextualization of pre-existing material. The Dadaists of the 1920s, notably Tristan Tzara with his ``cut-up'' technique, and later the Beat Generation poets in the 1950s, such as Brion Gysin and William S. Burroughs \cite{adema2017cut}, significantly contributed to the development and popularization of found poetry. They explored the potential of random and aleatory processes in literary creation, laying the groundwork for contemporary practices such as Magnetic Poetry.

The effect of constraints on creative thinking is illustrated in the practices of found and Magnetic Poetry. By imposing limitations on the selection and arrangement of words, these poetic forms paradoxically liberate the creative process. This phenomenon underscores the role of constraints not as creative barriers, but as catalysts that stimulate creative thinking and problem-solving. Constraints serve a crucial role in the creative process by focusing attention and reducing the overwhelming possibilities that can lead to creative block \cite{medeiros2014not}. Working within a fixed lexicon, participants are naturally led to forge unexpected connections between words and ideas. 

\section{System Design}
\label{s:system_design}
In this section we discuss the system design and overall concept of the \MP, beginning with the design rationale for using magnetic poetry as an interface.

\subsection{A Poetic Interface}
\label{ss:poetic_interface}

To speculatively explore possibilities for human-AI interaction we opted to use poetics as the means of communication between human and machine. By using poetry or poetic language, we mask or obscure the authorial intention of the human and AI authors, placing both on a more equal footing. While poetry and poetic text are the primary communication mechanism for the system, its overall goal is to support creative ideation and divergent thinking rather than being an AI poet. As such, responses are more often \emph{poetic} rather than \emph{poetry}, for example taking the form of aphorisms, metaphor or allegorical narrative rather than literally being poems.

\begin{figure*}
    \centering
    \includegraphics[width=0.8\textwidth]{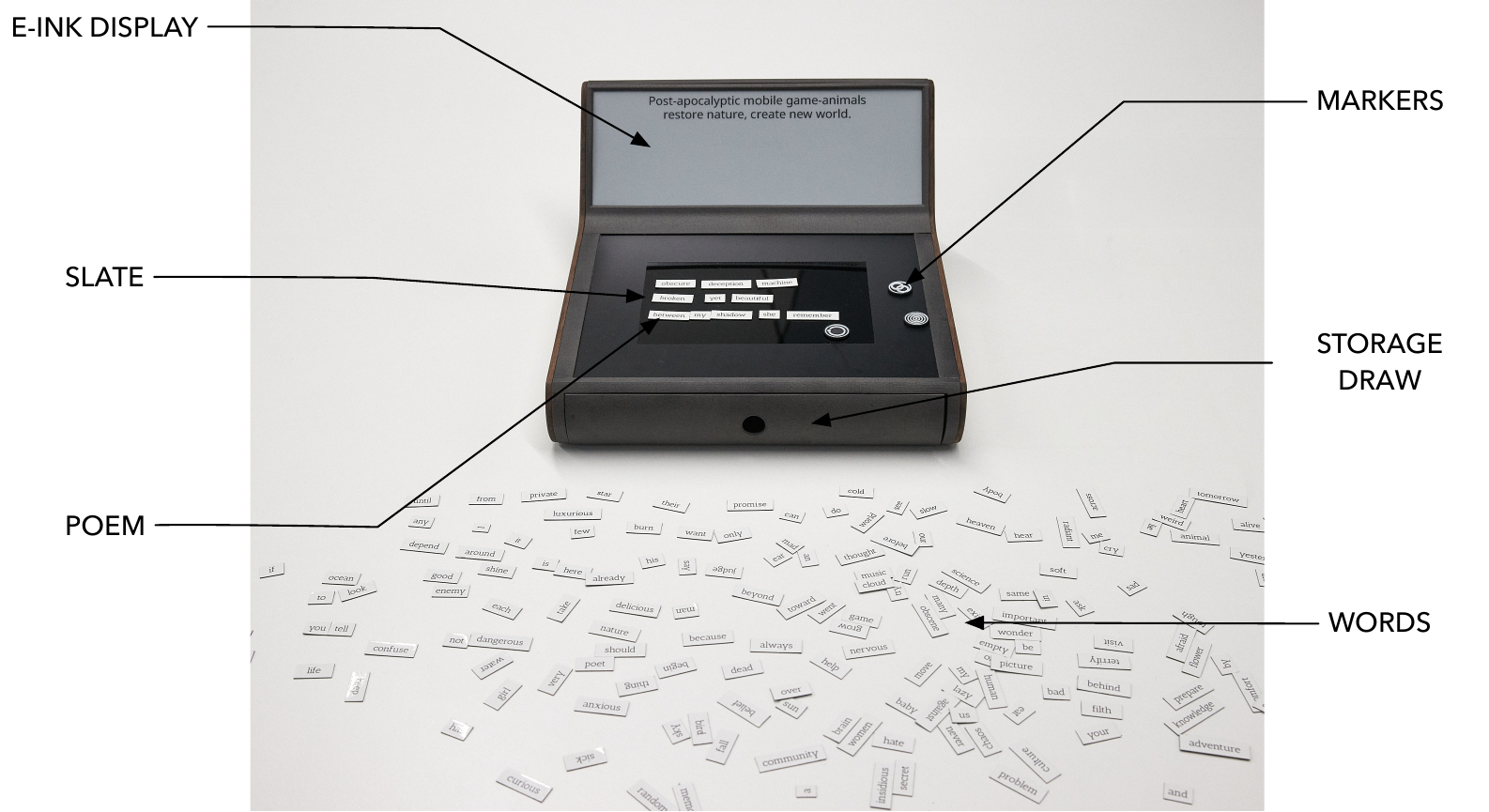}
    \caption{The Mimetic Poet machine, showing key elements of the device}
    \label{fig:mp}
\end{figure*}

The use of magnetic poetry lowers the ``barrier of entry'' for people who may not be used to creating poetic text and may not consider themselves poets or even authors in any traditional sense. As both an ``interface'' and form of expression it has a number of advantages, which we classify into three categories:
\begin{description}
     \item[constraints:] magnetic poetry uses a limited vocabulary of possible words from which to compose the poem. This helps to reduce decision paralysis, even picking up words at random can be used to begin. The constrained size of available space helps keep poems brief and to the point.
    \item[usability:] the playful, engaging and physical manipulation of words is simple, moving or editing of words inspires thought on meaning and context.
    \item[physicality:] physical words allow for serendipitous creativity, e.g.~seeing interesting combinations of words in close physical proximity around the device (Fig.~\ref{fig:mp}) can inspire a poem's creation. Being in a shared physical space (rather than on a personal device such as a phone or laptop screen) encourages co-creation in groups of people, who can collaborate on a poem's construction.

\end{description}

\subsection{Slow technology}
\label{ss:slow_technology}
Standard ``chat'' style interfaces with LLMs take the form of a continuous dialogue where the person asks a question then receives an instantaneous response. While such dialogues are suited to interactions such as question and answer, or specific goal or task directed activities (``summarise the following report'', ``give me a recipe with these ingredients'', etc.), they reinforce concepts of immediacy and divisibility in task-focused language.

In contrast, the use of a physical interface and its affordances emphasises a different approach to using the technology and how to engage with it. Having to manually assemble a poem from a fixed set of words as physical objects requires a distinct mode of thought over standard conversation or text messaging. Assembling the poem and waiting for a response allows for the human participant to consider how an AI might interpret their intention from the poem's text. The use of an e-ink display means the machine's responses will always be visible on screen (even when the power is turned off), giving a stronger sense of permanence and significance over traditional chat-based interfaces.

This aligns with the philosophy behind the slow technology approach \cite{odom2012slow,hallnas2001slow}, which advocates for more reflective human-machine interactions. This counters the fast-pacing, immediate nature of contemporary technology use, as exemplified by the question-response mode of the chat interface described above. By adopting this ``slow technology'' approach, we implicitly require the human participant to slow down in their interactions with the AI, leaving more space for contemplation, reflection, and evaluation of the exchanges, elevating their significance and prominence both physically and temporally. Our hypothesis is that, together, these features will encourage and support more nuanced understanding of an AI's role and capabilities, while at the same time assisting with human creative thinking.
    
In addition to expressing oneself to the AI through magnetic poetry, constraining the language model to reply in a poetic way helps to circumvent problems of both intention and hallucination -- there is no ``wrong'' way to write a poem and issues of truth or factual accuracy are less relevant than in more didactic or information-based tasks.

\subsection{The Device}
\label{ss:device}
\begin{figure}
    \centering
    \includegraphics[width=0.5\textwidth]{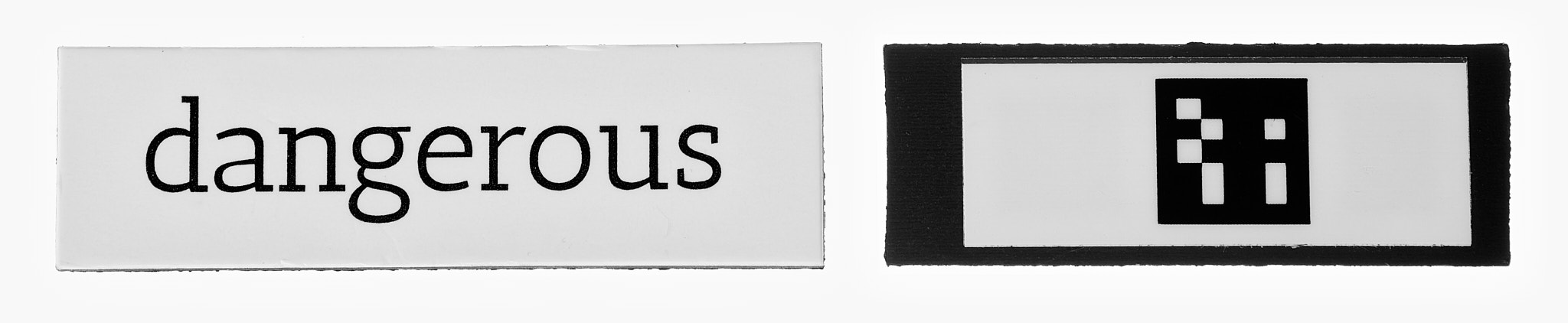}
    \caption{A magnetic poetry word (left) and the fiducial marker attached on the reverse side (right)}
    \label{fig:marker}
\end{figure}

The \MP~is designed as a stand-alone device that can sit easily in a studio, workplace or home (Fig.~\ref{fig:mp}). It consists of a flat surface (the \emph{slate}) upon which you can assemble your poem to be read by the system. A small drawer underneath the slate is used to store individual words used to create the poem. We found that placing the words around the device facilitated easy composition of poems. The words themselves were sourced from a standard magnetic poetry kit\footnote{The standard magnetic poetry kit has several hundred words, we used a subset of 175 words in the initial version of the system.}. On the rear side of each word is a small fiducial marker (Fig. \ref{fig:marker}) that a machine vision system can recognise when the word is placed on the slate  \cite{ROMERORAMIREZ201838}. We use this marker recognition as the mechanism for communication between human and machine.

\begin{figure}
    \centering
    \includegraphics[width=0.5\textwidth]{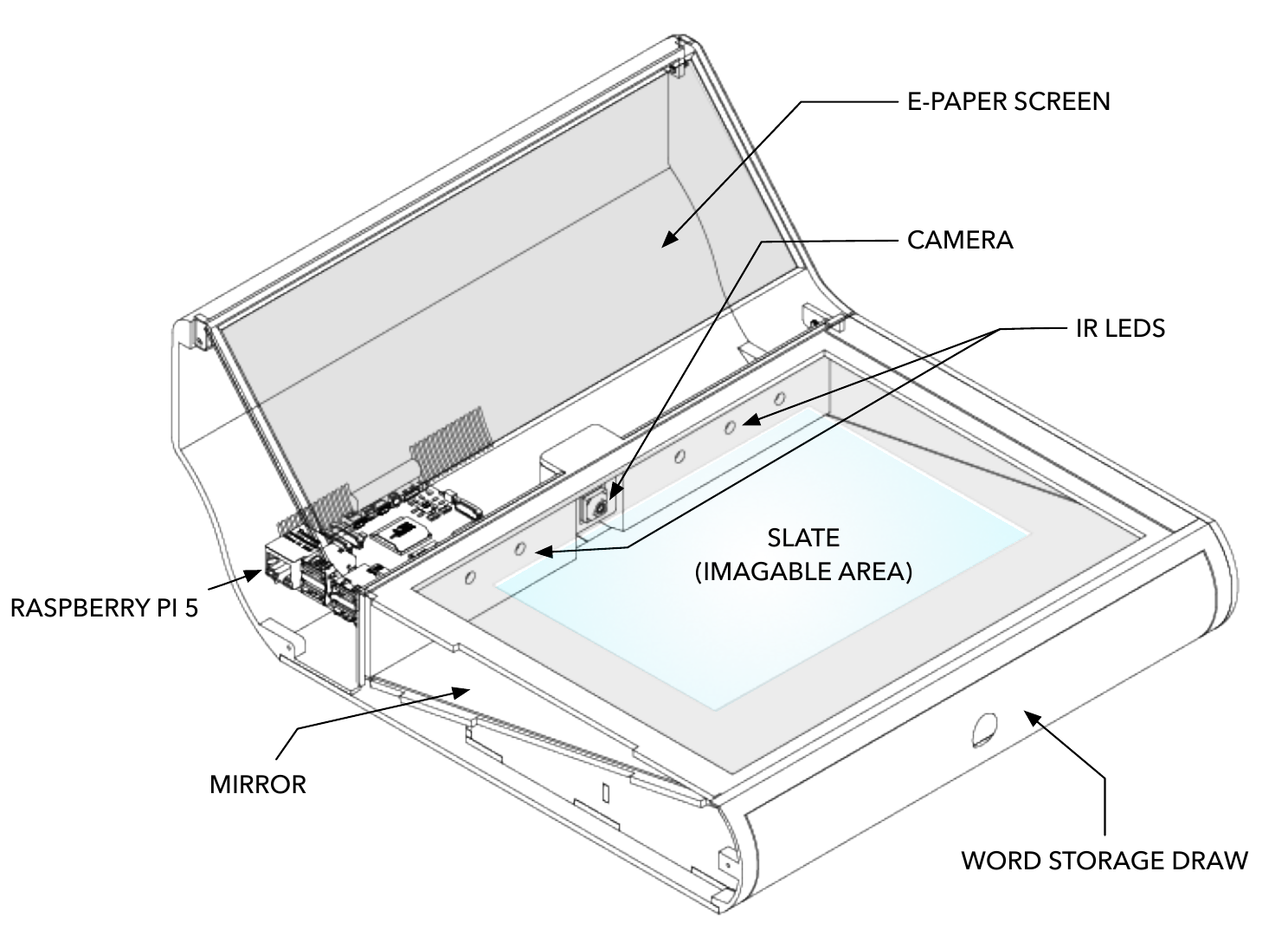}
    \caption{Schematic diagram of the Mimetic Poet Slate}
    \label{fig:schema}
\end{figure}

Inside the device, a camera and mirror system is used to detect the words placed on the slate. A Raspberry Pi 5 computer performs image processing and fiducial marker detection to identify the words placed on the slate in the sensing area (Figure \ref{fig:schema}). We use a camera that is sensitive to infrared (IR) light and apply a visible light filter to the lens, minimising any adverse effects of external visible light on the marker recognition and allowing the device to be used irrespective of external lighting conditions. An array of IR LEDs inside the device provides direct illumination of any objects placed on the slate.

\begin{figure}
    \centering
    \includegraphics[width=0.5\textwidth]{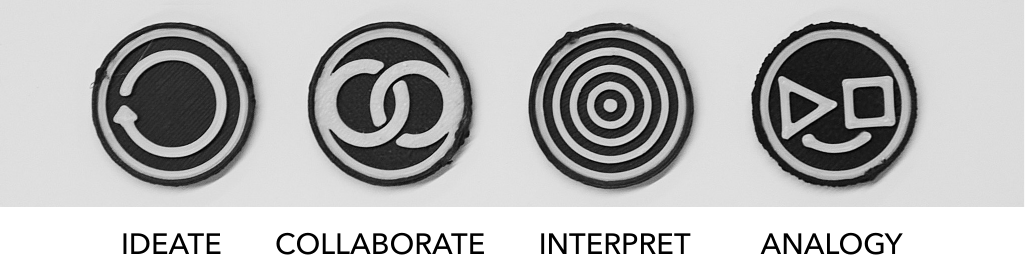}
    \caption{Four special markers that determine the poet's mode when placed on the slate.}
    \label{fig:modes}
\end{figure}

As words can be placed in an arbitrary location we developed an algorithm that captures the position and orientation of each word then uses a series of heuristics to determine how best to translate new lines and ensure correct word ordering. The aim was to match, as closely as possible, how a person would ``read'' the sequence of words placed on the slate. This was essential as we found participants often used interesting physical arrangements of the words as part of a poem's visual aesthetic.

To compute the correct ordering of placed words we developed a novel sorting algorithm that uses ray-tracing \cite{Whitted1980} to determine the best sequencing of placed markers. The marker detection system provides the positions of the centre ($m_c$) and the four corners of each detected marker. The marker with the highest vertical ($y$) position is first selected and removed from the list of unsorted detected markers. The top-left and bottom-left corners of the selected marker are subtracted from each other to form a vector representing the left edge of the marker. Next, we compute the tangent vector to the marker edge vector, $\vec{v}_t$, which is used to create a line, $l$ with start and end points $k$ units to the left and right of the maker respectively. 

\begin{equation}
  \begin{split}
    start & = k \vec{v}_t + m_{c}\\
    end & = -k \vec{v}_t + m_{c}\\
    l & = end - start
  \end{split}
\end{equation}

As the units used by the system are measured in pixels, we set $k = 1000$.
This line is then intersection tested \cite{Eberly2001} across each remaining unsorted marker, using a circle to represent the bounds of the marker.
\begin{equation}
    p_{c} = start + \left((m_{c} - start) \cdot \frac{l^2}{\left | l \right |^{^{2}}}\right) 
\end{equation}
Markers that collide with the line and the initial marker that was used to create it are removed from the unsorted list and moved to a new list that represents a single line of tiles:

\begin{algorithmic}
    \If{$\left | p_{c} - m_{c} \right | < tileHeight$}
         \State move $m$ to \textit{line list}
    \EndIf
\end{algorithmic}
\vspace{0.1cm}

\noindent This list is then sorted by the distance of each marker from the start point of the line. This ensures that even if the highest marker wasn't the left-most tile in the line that the order is correct. It also means that the sorting algorithm works with significantly skewed/diagonal lines of tiles and even upside down tile will sort in a meaningful way (i.e. they will be read right to left because the start and end points of the line will have flipped). After all the markers intersecting the line have been processed the algorithm is repeated until the unsorted marker list is empty.


If no word is moved for a few seconds, the text of the poem, including new lines, punctuation, etc. is converted to a string and the string sent to the AI subsystem (described in detail below).
In addition to placing words on the slate, we designed four special markers (Fig.~\ref{fig:modes}) that allow participants to select a different mode or personality for the AI.

We initially experimented with the idea that the markers should indicate the kind of response the participant is looking for from their poem:
\begin{enumerate}
    \item \textbf{interpret}: the system attempts to interpret the input poem, based on emotional tone or content, and provide a ``reading'' of the participants mental or emotional state.
    \item \textbf{collaborate}: the system tries to collaborate with the participant on poetry generation, generating a variant on the text supplied using the same words as are available to the participant.
    \item \textbf{ideate}: the system uses the input poem to help with ideation, responding with an idea or strategy that builds upon similarities between the poetic concept presented and ideas that the author may be interested in.
    \item \textbf{analogy}: the system constructs an analogous text based on concepts from a different discipline.
\end{enumerate}

In early testing with this scheme, we observed that while participants liked the ability to control and direct the system, the responses were often wordy or too cryptic for people to see obvious relationships between poem and response. Moreover, differentiating the modes, both visually in terms of the marker icon and conceptually in terms of the response, proved difficult. To address these issues, we developed a prompt chaining scheme \cite{Wu2022} to better control the LLM's responses to the input poems. 

\subsection{Prompting the LLM}
\label{ss:prompting}

The design of prompts used internally plays a pivotal role in steering the direction and quality of the outputs generated by the system. We implemented a set of prompt chains for the preliminary testing of the system. To this end, we utilised LangChain \cite{Chase_LangChain_2022}, a python package which facilitates prompt design and chaining with interchangeable LLM services. For this study, we used OpenAI's GPT4, accessed via the Python API. 

The prompts were written with respect to the four modes (Figure \ref{fig:modes}) and are shown in Table \ref{tab:promptchain}. For each mode, we first constructed a prompt that exemplified the desired outcome of the mode. The second stage of the prompt chain was to summarise the first response, placing a restriction on the LLMs output. Incorporating a second stage in the prompt chain, where the system's output is condensed or reinterpreted, introduces an essential feedback mechanism. This step is not only a constraint, but filters the initial response, promoting clarity, brevity, and a different presentation of the original output. This was thought to generate unusual, poetic, and particularly cryptic responses. This configuration serves as an example from which participants of the study (detailed below) based their initial impressions, and suggested modifications or improvements to the prompts. 

\begin{table*}[htbp]
\centering
\begin{tabular}{>{\raggedright\arraybackslash}p{2cm} >{\raggedright\arraybackslash}p{8cm} >{\raggedright\arraybackslash}p{6cm}}
\toprule
\textbf{Mode} & \textbf{Prompt 1} & \textbf{Prompt 2} \\
\midrule
Interpret & I just wrote the following text: \{poem\}. Speculate on what I'm feeling when writing this. Please keep the interpretation short (2-3 sentences). & Summarise this: \{response\} in only 5-15 words. \\
\addlinespace
Collaborate & Select words from the following text: \{poem\} to form a question that the text seems to be asking or addressing. Then, use other words from the text to answer it (2-3 sentences). & Summarise this: \{response\} in only 5-15 words. \\
\addlinespace
Ideate & The user just input the following text: \{poem\} Try and develop a creative idea or strategy that builds upon similarities between these words/concepts presented. Please keep your response short (2-3 sentences). & Reword your answer here: \{response\} in only 5-15 words. \\
\addlinespace
Analogy & Reframe this the following text with reference to a different discipline: \{poem\} & Repeat the following: \{response\} except obscure it further. \\
\bottomrule
\end{tabular}
\caption{The prompt chaining for each mode, with `{poem}' variable being the input poem by the user, and `{response}' being the LLM output resulting from the first prompt.}
\label{tab:promptchain}
\end{table*}

\section{Study}
\label{s:study}

To better understand how the efficacy of the \MP~as both a human-AI interface and system for supporting creative thinking, we undertook a preliminary study within our research laboratory. We placed the device in a communal area of the workplace where people often gathered. Participants ($N=14$) were instructed on the basic purpose of the device (a machine to support creativity that uses magnetic poetry to communicate with an AI) and encouraged to use it as they saw fit. The device was installed over a two week period and during that time participants generated 413 individual poems (an average of 29 poems/participant). No remuneration was provided for participation and the project was approved by our university Ethics committee.

The most popular mode was ``collaborate'' (59\%), followed by ``ideate'' (19\%), ``interpret'' (14\%) and ``analogy'' (8\%). Participants used only 111 of the 175 available words, with the AI responses generating 6,833 unique words, a forty-fold increase. Interestingly, the most popular human-selected words (\textit{human, dead, deception, memory, machine, bad, filth, heaven, delicious, eat}) were almost identical to those selected by the AI (\textit{human, memory, community, machine, flower, heaven, filth, wonder, empty, nature}), indicating that the responses often used words from the input text.

At the conclusion of the study period, we held an in-person focus group to hear participants' thoughts and experience of using the device. The focus group consisted of semi-structured interviews and open discussion in the same area that the \MP~was installed, allowing participants to directly demonstrate their use of the device and to refer to specific aspects of its design by indicating or showing. The focus group was attended by participants ($N=6$). All comments were audio recorded and transcripts produced, along with researcher's notes and still photography.

\begin{figure}[ht]
    \centering
    \includegraphics[width=0.5\textwidth]{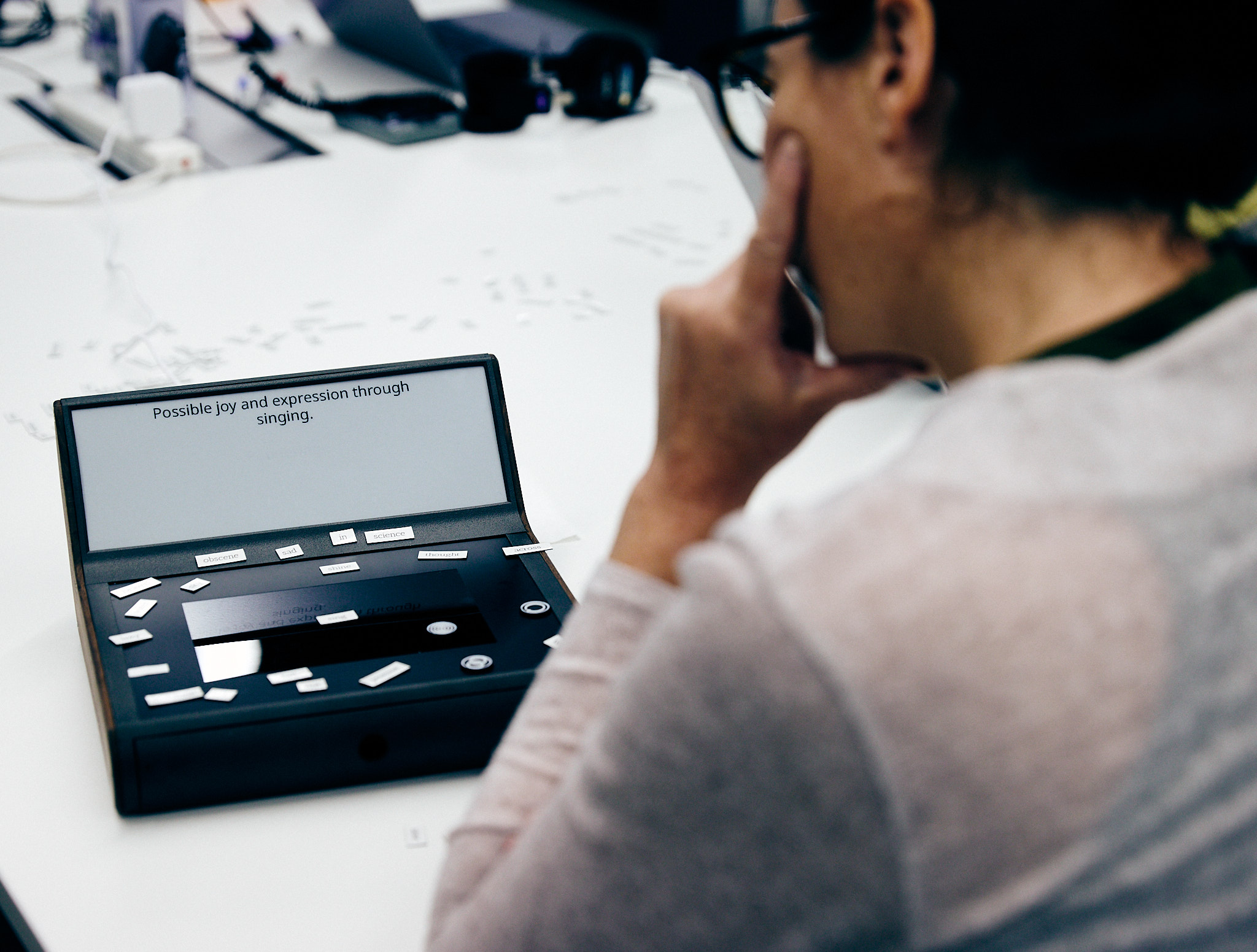}
    \caption{A participant exploring the \MP~during the focus group session.}
    \label{fig:focus}
\end{figure}

\subsection{Methodology}
We collated the transcripts, notes and internal machine logs and undertook a basic analysis using two methods. Firstly, individual researchers analysed the text and notes and drew out specific themes of importance from participants responses. The research team then came together to compare findings and, following discussion, individual findings were merged, resulting in four identifiable themes: (i) the interface itself and physicality of interaction; (ii) usability; (iii) Participant's overall perceptions of the AI; (iv) The AI's responses to participant's poems. 
We discuss findings under each of the themes below.

\subsection{Themes}

\begin{quote}
    ``I have really enjoyed being invited to engage in something playful. It is a pleasant break from work that puts me in a good mood and perhaps relieves work related `stuckness' or frustration. I think being playful is important for creative people.''
    \flushright --- Focus Group Participant
\end{quote}

\subsubsection{Interface/Physicality}

Participants of our focus group found the interface ``playful'' and ``like a game'', with the magnet poetry component of the system intuitive and fun to engage with. One participant remarked that the device was ``hard not to interact with''. A majority of participants commented on the physical design being ``retro'' and reminiscent of old or obsolete technologies. This led to a good deal of curiosity and interest. Other participants also enjoyed the fact that it ``free[s] your mind from the urgency of working with a screen''. The choice to embed the technology into a physical device was enjoyed by all participants, and reminded them of a ``board game'' rather than carrying the usual connotations of interacting with AI.

With respect to the limited vocabulary available, participants found this both enjoyable and frustrating; ``that's the beauty of fridge poetry''. However, with reference to the machine's responses, one participant noted it annoying that the system has, in contrast, an unlimited vocabulary: ``all I have is simple nouns and verbs, and it comes with high philosophical concepts''. Some participants enjoyed the process of scanning through and picking up the individual word magnets as an ``inspirational'' activity. Yet, other participants commented on their preference to be able to write freely to the system. For example, open writing interface would be ``giving more space for the participant to explore \ldots
rather than putting some random words that someone else picked''. While the constraint made it easier for some to engage, there is a sense that some participants struggled to communicate with the AI; that their individuality was lost in the process.

\subsubsection{Usability}
All participants found the device easy and intuitive to use and were impressed with the reliability of the word recognition. However some commented on the delay between finishing the poem and getting a response being too long. This expectation for immediate responses contrasts the slow-technology approach adopted in the \MP, and while participants found this to be an issue in their initial interactions, they also commented that this would not be a problem for sustained use ``I would definitely have something like this if I'm working continuously on a writing piece, if I need to ignite some ideas'', where it could be seen more ``as a companion'', in contrast  of it being ``a one encounter'' type of interaction. Upon further discussion participants suggested wanting more immediate feedback from the device in terms of what it is doing ``it takes time to figure out that you need to wait and that there could be some glitches in between''. 

Overall, participants identified a personal or individual use as their preferred mode of interaction ``I feel like it could be a private moment, and then you can share it with some other people'', ``just for me, for my reflective practice'', ``I would love to have one of these at home to just play around and see what happens''. However, they also highlighted the benefits of more social or group use, ``it could even be fun to use it in a session like this when we're trying to ideate things together'', and even one participant commented how they got inspired on their own work seeing the response another participant got from the device ``but it wasn't something that I did. It was something that \ldots was doing before''.

Finally, participants also mentioned the ability to customise their device as something they would find useful ``as time goes by, it does have more information about you \ldots it would introduce dynamic''. Additionally, although they found it fun to use, some of the participants mentioned getting tired of the interactions over a longer period. Customisation as the device learns from human interactions might help alleviate this issue. 

\subsubsection{AI Perception}
Participants pointed out that the responses from the AI have a very similar positive tone, which was a ``limiting factor’’ for creative expression, ``if you are limited to just that spectrum of the emotion \ldots then you are never going to really explore things properly.’’ As tech companies try to make LLMs safe for public use, they have adopted a dominant positive tone in the models responses. In a sustained interaction, this may lead to a gap between someone's intentions and the device's ability to meet those intentions.

We acknowledge that controls for this must be in place, but for creative expression a balance may be required. On this, participants mentioned they would like the ability to change the ``personality'' of the device ``I think when it displays personality it is engaging and enjoyable'', citing examples when the responses seemed to have ``some kind of hidden sentiment, like sarcasm or double meaning'' as more engaging. Although participants did not report this as a factor to enhance creative thinking, they repeatedly refer to this type of interaction as playful and enjoyable, something that would set them in a better mood, which previous studies have shown to enhance creative thinking \cite{vosburg1998effects,baas2008meta}. 

\begin{table*}[htbp]
\centering
\begin{tabular}{>{\raggedright\arraybackslash}p{2cm} >{\raggedright\arraybackslash}p{4.5cm} >{\raggedright\arraybackslash}p{6.5cm}}
\toprule
\textbf{Mode} & \textbf{Input Poem} & \textbf{Output Text} \\
\midrule
Interpret & brain problem\newline see over here\newline each bad judge\newline secret life insidious\newline their obscene picture\newline is already across from\newline a good few &  You seem to be grappling with feelings of frustration and anxiety, possibly due to feeling misunderstood or judged. \\
\addlinespace
Collaborate & hate delicious body\newline beautiful anxious heart & Delicious hate, body beautiful,\newline Anxious heart, artfully dutiful. \\
\addlinespace
Ideate & do promise a radiant world\newline forest see animal\newline their same cry\newline beyond science\newline slow broken heaven & Post-apocalyptic mobile game-animals restore nature, create new world \\
\addlinespace
Analogy & shine promise water\newline thought until flower\newline already soft & Dewy water softly caresses the budding flower, refreshing the mind's hardened soil \\
\bottomrule
\end{tabular}
\caption{Example text generated by the system for each mode, based from participants input poems}
\label{fig:examples}
\end{table*}

\subsubsection{AI Response}
In addition to wanting responses with different tones or as the participants put it, ``personalities'', they also mentioned that the more poetic-like responses were the most interesting as they differed from the usual descriptive LLM style. On this, most of the participants agreed that the {\em Analogy} mode took them closer to the kind of preferred response, with one of them saying ``is the most proactive answer \ldots it feels the closest to poetry''. 

An interesting insight was also on how much detail to give about the nature of a response, with some participants stating that they would like to understand ``the intent of the prompt'' as this would help them ``understand what's happening behind it''. These participants believed this could lead to the ability to ``tune [the device] to your own particular needs''. However, other participants stated that they would prefer not knowing since they could assign their own interpretation to the response ``me putting myself in, could dictate what I'm going to get out [which] is the whole purpose''. 

\section{Discussion}
As the system designers, we found the focus group feedback very helpful in considering how the current design is perceived by its users, and ways we can improve it. However, as a small study on an initial prototype, our work has a number of limitations. Only a subset (6) of participants took part in the focus group, so we weren't able to canvas all participant's opinions of the device. However, the large number of interactions logged over the two week evaluation period suggests that the \MP~received a lot of attention and use -- an average of 41 new poems were created each day. Secondly, participants were members of our laboratory and were all creative practitioners or researchers highly familiar with technology and the creative use of Artificial Intelligence. This is not necessarily a disadvantage, as all participants had previously attempted to use tools like ChatGPT for creative ideation or to support creative thinking. The majority found such tools ill-suited to the task and saw the \MP~as a playful alternative. 

Although at the heart of our design is the idea of encouraging reflection following the principles of slow technology, the study showed us that this will represent a challenge as some participants longed for the immediacy of current interactions. Research in the design of these types of technologies, which are intended to ``surround us and therefore 	is part of our activities for long periods of time’’ \cite[pp 161]{hallnas2001slow}, have emphasised the need of persistent use in order to develop such intended relationship \cite{maze2005form}. For this, a study with a more sustained period of use is needed.

Lastly, while a number of participants reported finding using the tool stimulating and interesting, we did not find that the device, in its current form, was directly responsible for solving practical creative problems.

Despite these limitations, we feel that the \MP\, has significant potential, which we hope to realise in the next iteration, currently under development. The feedback from our study has suggested several design changes that would improve engagement and quality of responses. These include:
\begin{description}
    \item[Interface:] additional feedback while composing a poem was seen as favourable amongst several participants; adding markers that represent different AI ``personalities'' that allow participants to customise the responses as an alternative to the current mode markers (Fig.~\ref{fig:modes}).
    \item[AI responses:] we currently use a simple prompt chaining technique (Table~\ref{tab:promptchain}) to prompt GPT-4. Further development of more advanced prompt chains is needed to increase the quality and suitability of the AI responses. Experimenting with other LLMs, in particular ones that can be easily fine-tuned (such as Mistral7b) would help overcome the overly positive and didactic responses participants often received from the current system.
    \item[AI perception:] currently our zero-shot prompting does not allow the LLM to make use of past interactions, leading to a more transactional interaction from participants. Adding the ability to personalise the model and incorporate past interactions into the prompt chain would allow the LLM to also reference a participant's history of interactions, enriching the experience over the long term.
\end{description}

\section{Conclusions}
In this paper have presented the \MP, a novel device that uses magnetic poetry as the means of communication with a LLM for the purposes of encouraging creative thinking and ideation. We designed and built the device, then evaluated it with a group of participants over a two week period, canvasing views of user experience in a focus group session. Our findings showed that some encounters with the device helped participants in creative thinking and ideation, and that the interface was a desirable alternative to traditional chat-based interfaces, with our interface preferred as an inspirational device.

We view these alternative human-AI interaction methods as a means to catalyse wider conversations about AI's role in human creativity, presenting new ways for us to engage with artificial systems. By examining logs of human input and AI responses, it's clear that humans are currently the more creative of the two participants\ldots for now.

\bibliographystyle{iccc}
\bibliography{extracted}

\end{document}